\documentclass[twocolumn,prl,superscriptaddress,showpacs,floats]{revtex4}
\usepackage[pdftex]{graphicx}
\usepackage{dcolumn}
\usepackage{bm}
\begin{document}

\newcommand{\LSMO}{La$_{0.5}$Sr$_{1.5}$MnO$_{4}$}

\title{Orbital order in La$_{\bf 0.5}$Sr$_{\bf 1.5}$MnO$_{\bf 4}$: 
beyond a common local Jahn-Teller picture}

\author{Hua Wu}
\thanks{Corresponding author; wu@ph2.uni-koeln.de}
\affiliation{II. Physikalisches Institut, Universit\"{a}t zu K\"{o}ln,
Z\"{u}lpicher Str. 77, 50937 K\"{o}ln, Germany}
\affiliation{Department of Physics, Fudan University, Shanghai 200433, China}
\author{C.~F.~Chang}
\affiliation{II. Physikalisches Institut, Universit\"{a}t zu K\"{o}ln,
Z\"{u}lpicher Str. 77, 50937 K\"{o}ln, Germany}
\author{O.~Schumann}
\affiliation{II. Physikalisches Institut, Universit\"{a}t zu K\"{o}ln,
Z\"{u}lpicher Str. 77, 50937 K\"{o}ln, Germany}
\author{Z.~Hu}
\affiliation{II. Physikalisches Institut, Universit\"{a}t zu K\"{o}ln,
Z\"{u}lpicher Str. 77, 50937 K\"{o}ln, Germany}
\affiliation{Max Planck Institute for Chemical Physics of Solids, 
N\"othnitzerstr. 40, 01187 Dresden, Germany} 
\author{J.~C.~Cezar}
\affiliation{European Synchrotron Radiation Facility, Bo\^{i}te Postale 220, 
38043 Grenoble C\'{e}dex, France}
\author{T.~Burnus}
\affiliation{II. Physikalisches Institut, Universit\"{a}t zu K\"{o}ln,
Z\"{u}lpicher Str. 77, 50937 K\"{o}ln, Germany}
\author{N.~Hollmann}
\affiliation{II. Physikalisches Institut, Universit\"{a}t zu K\"{o}ln,
Z\"{u}lpicher Str. 77, 50937 K\"{o}ln, Germany}
\author{N.~B.~Brookes}
\affiliation{European Synchrotron Radiation Facility, Bo\^{i}te Postale 220,
38043 Grenoble C\'{e}dex, France}
\author{A.~Tanaka}
\affiliation{Department of Quantum Matter, ADSM, Hiroshima University, 
Higashi-Hiroshima 739-8530, Japan}
\author{M.~Braden}
\affiliation{II. Physikalisches Institut, Universit\"{a}t zu K\"{o}ln,
Z\"{u}lpicher Str. 77, 50937 K\"{o}ln, Germany}
\author{L.~H.~Tjeng}
\affiliation{II. Physikalisches Institut, Universit\"{a}t zu K\"{o}ln,
Z\"{u}lpicher Str. 77, 50937 K\"{o}ln, Germany}
\affiliation{Max Planck Institute for Chemical Physics of Solids,
N\"othnitzerstr. 40, 01187 Dresden, Germany}
\author{D.~I.~Khomskii}
\affiliation{II. Physikalisches Institut, Universit\"{a}t zu K\"{o}ln,
Z\"{u}lpicher Str. 77, 50937 K\"{o}ln, Germany}


\begin{abstract}

The standard way to find the orbital occupation of Jahn-Teller (JT) ions
is to use structural data, with the assumption of a one-to-one 
correspondence between the orbital occupation and the associated JT distortion, $e.g.$ 
in O$_6$ octahedron. We show, however, that this approach in principle does not work 
for layered systems. Specifically, using the layered manganite La$_{0.5}$Sr$_{1.5}$MnO$_{4}$ 
as an example, we found from our x-ray absorption measurements and theoretical 
calculations, that the type of orbital ordering strongly contradicts the standard local 
distortion approach for the Mn$^{3+}$O$_6$ octahedra, and that the generally 
ignored long-range crystal field effect and anisotropic hopping integrals are actually 
crucial to determine the orbital occupation. Our findings may open 
a pathway to control of the orbital state in multilayer systems and thus of
their physical properties. 

\end{abstract}


\maketitle

Rich physical properties of $3d$ transition metal (TM)
oxides are largely related to an interplay among charge, orbital,
spin and lattice degrees of
freedom~\cite{Kugel-Khomskii-SOV,Tokura-Nagaosa-orbital}. In
particular, the orbital pattern plays a crucial role in magnetic
exchange~\cite{Goodenough-book} and in many other properties.
Therefore large attention is nowadays devoted to establishing the
detailed type of orbital order (OO) in oxides having degenerate orbitals, 
and different
methods can be applied. The oldest but simple and still the most
widely used one is based on the assumption that there is a
one-to-one correspondence between the orbital occupation and the
local distortion of the nearest-neighboring anion cage around
the TM ion, i.e. the local Jahn-Teller (JT) distortion.

For the doubly-degenerate case of $e_g$ electrons one can
characterize orbital occupation by the ``mixing angle'' $\theta$,
$|\theta\rangle=$ cos($\theta$/2)$|3z{^2}-r{^2}\rangle$ +
sin($\theta$/2)$|x{^2}-y{^2}\rangle$. Local deformations of the
oxygen octahedron around the TM ion can also be presented as a
superposition of the tetragonal distortion $Q_3$ and the
orthorhombic distortion $Q_2$, so that the general deformation of
the octahedra is $\theta'$,
$|\theta'\rangle=$ cos($\theta'$)$|Q_3\rangle$ +
sin($\theta'$)$|Q_2\rangle$. When using the JT mechanism of electron-lattice
interaction, the standard assumption, always made,
is that the ``mixing angles'' $\theta$ and $\theta'$ are the
same, and that the orbital mixing angle $\theta$ is determined by
the relation~\cite{Goodenough-book} $\tan(\theta)=$
$\frac{\sqrt{3}(l-s)}{2m-l-s}$, where $l$, $m$ and $s$ are the
long, middle and short distances, respectively, from TM to the 
nearest-neighboring 
ligands, all representing a local JT distortion. This rule is usually true
for isolated JT centers, and for cubic oxide materials where an additional
superexchange mechanism may also get involved in determining
an orbital state~\cite{Kugel-Khomskii-SOV,Oles2005}.
Even in the latter case, however, after such an orbital state is established,
a local distortion will in principle adjust to the orbital occupation,
such that in effect again the local distortion and the orbital occupation
would be the same.
But, as we will show below, this rule breaks down for certain layered systems,
because not only local distortions but also
the anisotropic long-range
contribution to the crystal field and the electronic kinetic (band) energy play
crucial role in determining the orbital state.

The half-doped single-layer manganite {\LSMO} turns out to be a
crucial testing and battling ground for the modeling of the
relationship between crystal structure, local electronic structure
and the magnetic properties. On the basis of crystal structure
data, resonant scattering measurements and theoretical
calculations~\cite{Huang_PRL_2004,Wilkins_PRB_2005,Stojic_PRB_2005,
Ebata_PRB_2005,Zeng_PRB_2008,Okuyama_PRB_2009},
the claim was made that the orbital pattern in the charge-ordered
(CO) state of this material with the CE-type antiferromagnetic
structure involves the cross-like $x^2$--$z^2$/$y^2$--$z^2$ type
orbitals for the Mn$^{3+}$ ions. The results seem to fit to the
standard local distortion model: local structural data show that
the O$_6$ octahedra around the Mn$^{3+}$ sites are not elongated,
but rather locally compressed, with four long and two short Mn-O
distances, thus `justifying' the cross-like type of orbital
occupation. Yet this finding is quite surprising in view of the
fact that earlier various theoretical 
studies~\cite{Mizokawa_PRB_1997,Solovyev_PRL_1999,Mahadevan_PRL_2001,
Daghofer_PRB_2006,Lee_PRL_2006}, not knowing the local structure
around the Mn ions, proposed the rod-like
$3x^2$--$r^2$/$3y^2$--$r^2$ type to explain the experimentally
observed magnetic structure. Also an alternative interpretation
\cite{Mirone_EPJB_2006} of the x-ray scattering data was not
considered to be persuasive.

We reinvestigated this question on high-quality single crystals of
{\LSMO}~\cite{Reutler_JCG_2003,Senff_PRB_2005,Senff_PRL_2006,Senff_PRB_2008},
using polarization dependent soft x-ray absorption spectroscopy
(XAS) at the Mn $L_{2,3}$ edges~\cite{Hossain_PRL_2008} and
multiplet cluster
calculations~\cite{DeGroot1994,Thole1997,Tanaka1994}. We also
performed density functional calculations within the
local-spin-density approximation (LSDA) and LSDA plus Hubbard $U$
(LSDA+$U$)~\cite{Anisimov_PRB_1993,Wien2k} with a full structural
optimization~\cite{Struct_opt}, thereby staying close to the
recently available structural
data~\cite{Zeng_PRB_2008,Okuyama_PRB_2009}.
All the experimental and computational details are given in the
Supplemental Material.

\begin{figure}
  \includegraphics[angle=0,width=6cm]{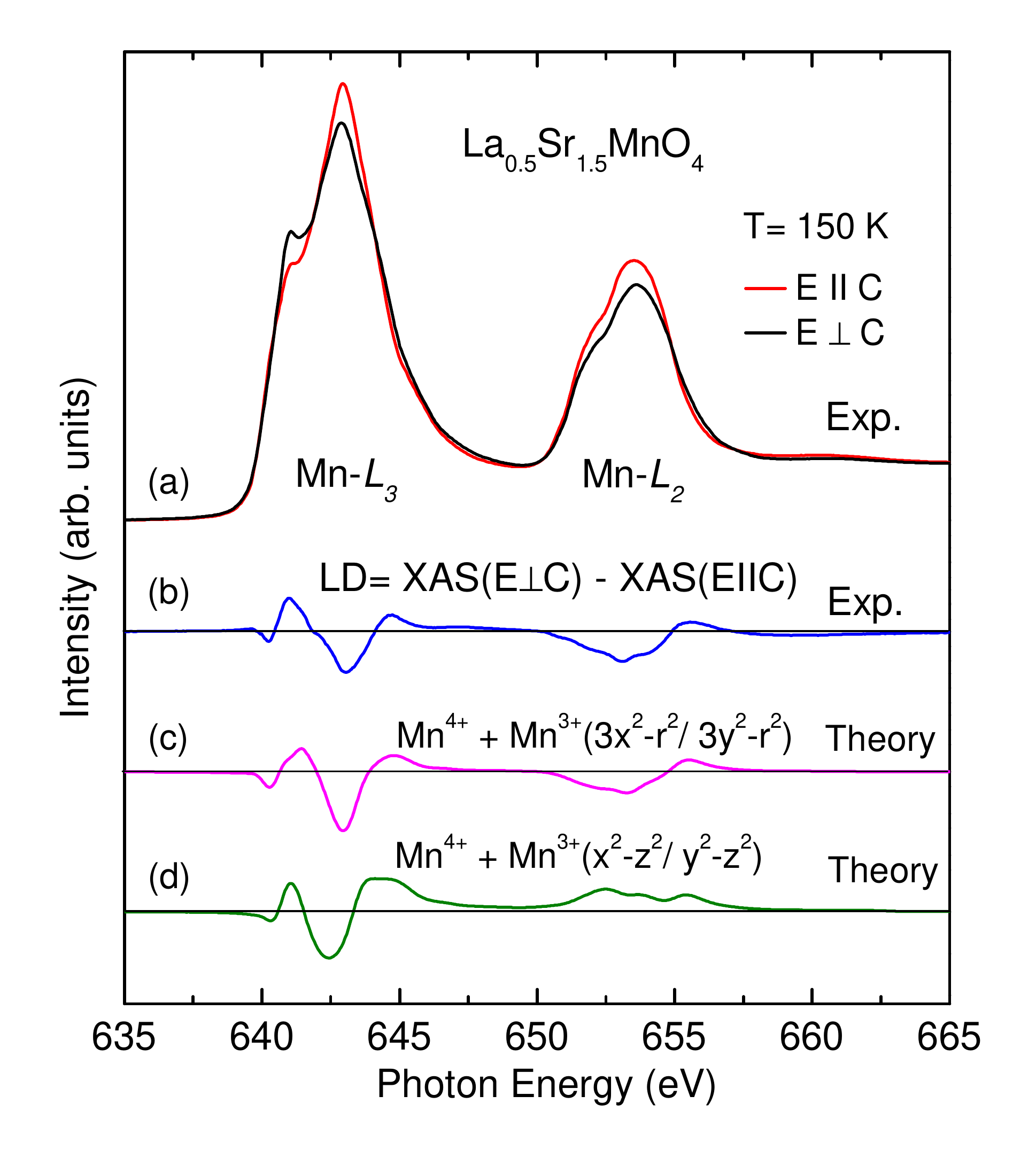}
  \caption{(Color online) Polarization-dependent Mn-$L_{2,3}$
    XAS spectra of La$_{0.5}$Sr$_{1.5}$MnO$_{4}$ for
    $\mathbf{E}\|\mathbf{c}$ (red curve) and $\mathbf{E}\bot\mathbf{c}$
    (black curve) taken at 150 K. (b) The corresponding
    linear dichroism (LD) spectrum.
    Theoretical LD calculated for orbital scenarios with
    (c) Mn$^{3+}$ $3x^2$--$r^2$/$3y^2$--$r^2$ or
    (d) Mn$^{3+}$ $x^{2}$--$z^{2}$/$y^{2}$--$z^{2}$ orbital occupation,
    each with the Mn$^{4+}$~$t_{2g}^{3}$.}
    \label{Fig-XAS-compare}
\vspace{-0.2cm}
\end{figure}

Figure~\ref{Fig-XAS-compare}~(a) shows our polarization dependent
Mn-$L_{2,3}$ XAS spectra of La$_{0.5}$Sr$_{1.5}$MnO$_{4}$ for
$\mathbf{E}\|\mathbf{c}$ (red curve) and
$\mathbf{E}\bot\mathbf{c}$ (black curve), where $\mathbf{E}$ is
the polarization vector of the light and $\mathbf{c}$ the
crystallographic axis perpendicular to the
$\mathbf{a}\mathbf{b}$-plane. The spectra were taken at 150 K,
i.e. below the orbital ordering temperature $T_{\rm OO}$ of 217 K.
Curve (b) depicts the corresponding linear dichroic (LD) spectrum,
defined as the difference between the two polarizations, i.e.
XAS($\mathbf{E}\bot\mathbf{c}$) -- XAS($\mathbf{E}\|\mathbf{c}$).
We note that our spectra are very different from those reported
earlier~\cite{Huang_PRL_2004,Merz_EPJB_2006}, and we will show
below that our spectra are truly representative for the
La$_{0.5}$Sr$_{1.5}$MnO$_{4}$ system.

To obtain some basic information concerning the orbital occupation
of the Mn$^{3+}$ ions, we can make use of a sum rule on the LD
spectrum~\cite{Csiszar_PRL_2005}. Integrating the LD spectrum
throughout the $L_{2,3}$ range, we find readily a negative net
value. Since the Mn$^{4+}$ ions have the half-filled $t_{2g}^{3}$
shell configuration, their contribution to the LD integral is
zero. The non-zero value of the integral is then due to the
Mn$^{3+}$ ions only. The negative value directly indicates that
the $e_g$ $holes$ have a more out-of-plane than in-plane character,
meaning that the $e_g$ electrons have their charge density more
in-plane than out-of-plane. This then favors the rod-like
$3x^2$--$r^2$/$3y^2$--$r^2$ type of orbital occupation and
directly rules out the $x^{2}$--$z^{2}$/$y^{2}$--$z^{2}$ scenario.

In order to further confirm the above experimental finding,
we have simulated the Mn-$L_{2,3}$ LD spectra using the
well-proven configuration interaction cluster model which includes
the full atomic multiplet theory~\cite{DeGroot1994,Thole1997,Tanaka1994}.
Figure~\ref{Fig-XAS-compare}(c) shows the calculated LD spectrum for the
Mn$^{3+}$ ions having the rod-like $3x^2$--$r^2$/$3y^2$--$r^2$
orbital occupation while the Mn$^{4+}$ ions are kept in
$t_{2g}^{3}$ configuration. We can see that all features of the
experimental LD spectrum are nicely reproduced, including the
negative value for the integral. As a check we also include curve
(d) which depicts the LD spectrum for the Mn$^{3+}$ ions with the
cross-like $x^{2}$--$z^{2}$/$y^{2}$--$z^{2}$ orbital occupation.
We notice significant discrepancies with the experimental
spectrum, especially at the $L_{2}$ edge where the sign of the
simulated LD is opposite to that of the experimental one.
Obviously, the integral value of this simulation (positive) has
also the wrong sign.

We would like to remark that the excellent agreement between the
simulation in curve (c) and the experimental LD spectrum can be
taken as evidence that we have been able to obtain spectra which
are representative for the La$_{0.5}$Sr$_{1.5}$MnO$_{4}$ system.
We have taken care that the high-quality single crystal was cleaved
\textit{in-situ} to obtain well ordered and clean sample area,
that charging problems were avoided by using a small sample area,
and that the polarization of the light rather than the sample was
rotated in order to keep the same sample area to be measured for
optimal comparison between the spectra taken with the two
polarizations.

Thus an important aspect that emerges directly from our
experimental LD and multiplet cluster calculations is that the
Mn$^{3+}$ ions have the rod-like $3x^2$--$r^2$/$3y^2$--$r^2$
orbital symmetry. This is contrary to a prediction based on
the local JT distortion:
from the very similar in-plane 2.01 {\AA~} ($x$) and the
out-of-plane 1.99 {\AA~} ($z$) Mn-O distances, and the shorter
in-plane 1.90 {\AA~} ($y$)~\cite{Struct_opt}, 
one could have expected
the $x^2$--$z^2$ level of  $e_g$ crystal-field to be lower. This
implies a mechanism for stabilizing the rod-like
$3x^2$--$r^2$/$3y^2$--$r^2$ OO in this layered
manganite different from the conventional local JT mechanism.

\begin{figure*}
 \includegraphics[angle=270,width=12cm]{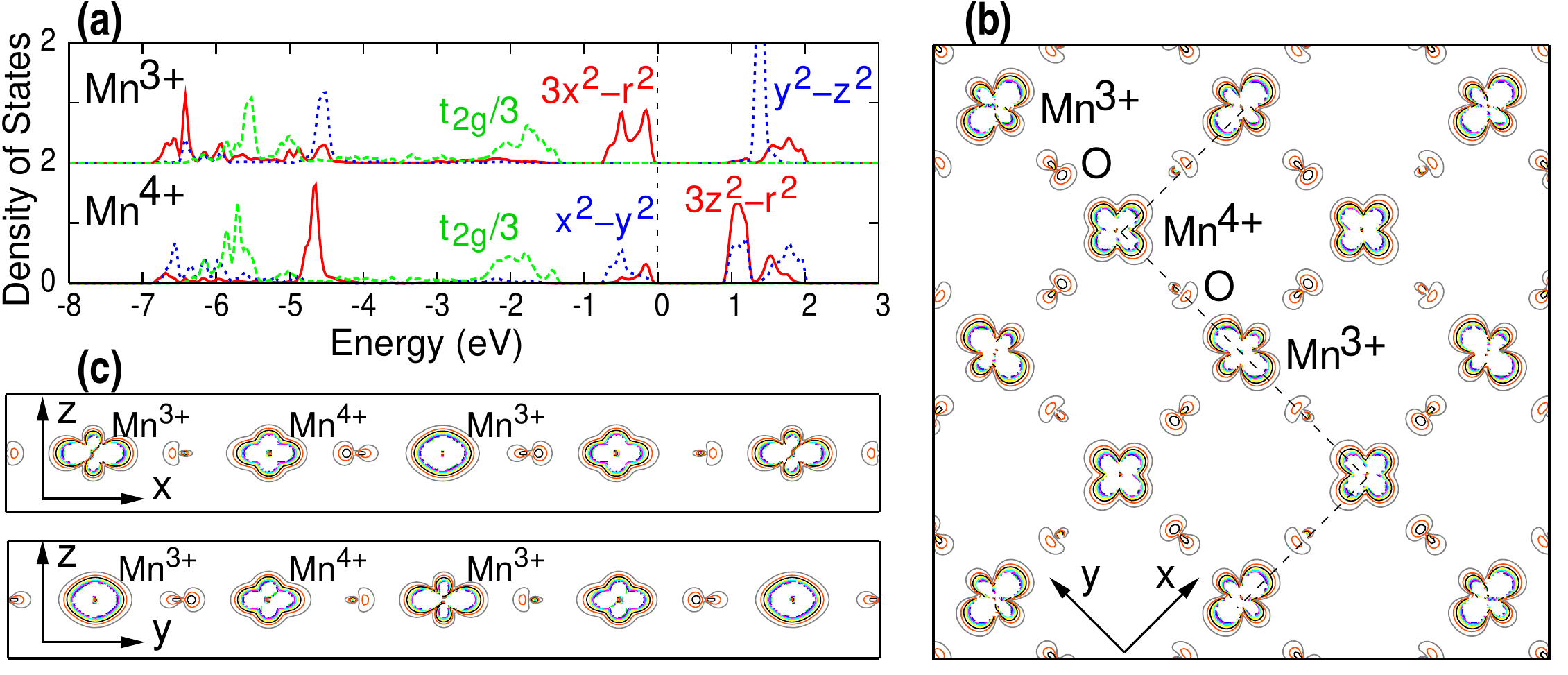}
  \caption{(Color online) (a) PDOS of the Mn$^{3+}$ and Mn$^{4+}$ ions 
in the CE-type antiferromagnetic state calculated by LSDA+$U$. (b) Charge
density contour plot (0.05-0.4 $e$/\AA$^3$) of the
Mn$^{3+}$-O-Mn$^{4+}$ network in the $xy$ plane for the occupied
$e_g$ states within 1 eV below Fermi level. The ferromagnetic
zigzag chain is marked by the dashed line. (c) Charge density
contour plot of the Mn$^{3+}$-O-Mn$^{4+}$ chain in the $xz$ and
$yz$ planes. (b) and (c) clearly show the occupied $3x^2$--$r^2$/$3y^2$--$r^2$
orbitals of the Mn$^{3+}$ ions.}
\label{Fig-DOS-Density} 
\vspace{-0.2cm}
\end{figure*}

To understand better this mechanism, we have studied the electronic and
magnetic structure of {\LSMO} and particularly the origin of
$3x^2$--$r^2$/$3y^2$--$r^2$ OO, using LSDA and
LSDA+$U$ calculations~\cite{Anisimov_PRB_1993,Wien2k} within the
space group $Bmbm$ which allows for the occurrence of the
experimentally observed CO and OO.
Figure~\ref{Fig-DOS-Density}(a) shows the orbitally resolved
partial density of states (PDOS) for the two inequivalent Mn sites,
calculated using LSDA+$U$ with $U$=5 eV and a Hund exchange energy
of 0.9 eV. An insulating gap of about 1.0 eV lies between the
split $e_g$ levels. For the Mn$^{3+}$ site, a pure $3x^2$--$r^2$
level appears just below the Fermi level. The
electron count is less than 1, i.e. 0.41$e$, due to the bonding
with the oxygens, as revealed by the mixed states below --4 eV. For
the Mn$^{4+}$ site, the amount of occupied $e_{g}$ states is only
a little less than for the Mn$^{3+}$, but it is highly mixed
$x^2$--$y^2$ and $3z^2$--$r^2$ hardly with any orbital polarization.
Note that in spite of the $e_g$-electron delocalization,
{\LSMO} can still be well categorized into a site-centered CO-OO
system; our calculations show that this state is more stable
than a bond-centered Mn-O-Mn polaronic state, by 70 meV/Mn.
This site-centered CO-OO state
for the half-doped case has also been confirmed by previous
experimental~\cite{Senff_PRL_2006} and
theoretical~\cite{Efremov2004} studies.

The difference between the Mn$^{3+}$ and Mn$^{4+}$ sites can also
be seen in Figs.~\ref{Fig-DOS-Density}(b) and
~\ref{Fig-DOS-Density}(c), where we show the $e_g$ charge
density contour plot on the $xy$ plane as well as on the $xz$ and
$yz$ planes of the Mn-O network. One can clearly observe the
$3x^2$--$r^2$/$3y^2$--$r^2$ OO at the Mn$^{3+}$ sites.
The Mn$^{4+}$ ions, on the other hand, have a nearly isotropic
$e_g$ charge density distribution. The CE-type ground state
magnetic structure can be understood via the
Goodenough-Kanamori-Anderson superexchange mechanisms, and this is
also confirmed by our calculations: the calculations of total-energy difference
between various magnetic structures allow us to derive that
the ferromagnetic coupling
within each zigzag chain is about --30 meV, and that the
inter-zigzag-chain antiferromagnetic coupling is much weaker,
being only about 2 meV. This agrees qualitatively with the
inelastic neutron scattering study~\cite{Senff_PRL_2006}.

It is important to note that the above results remain
qualitatively unchanged within a wide range of $U$ values: 0-8 eV.
The LSDA ($U$=0) calculation gives an insulating gap of 0.5 eV for
the CE-type magnetic structure and 0.32$e$ (0.05$e$) polarization
for the Mn$^{3+}$ (Mn$^{4+}$) $e_g$ orbitals. Switching on
the $U$, the gap increases up to 1.3 eV for $U$=8 eV and the
corresponding $e_g$ orbital polarization is 0.50$e$ (0.05$e$) for
the Mn$^{3+}$ (Mn$^{4+}$). The $3x^2$--$r^2$/$3y^2$--$r^2$ type of
OO is found to be $U$-independent.

\begin{figure}
 \includegraphics[angle=0,width=6.5cm]{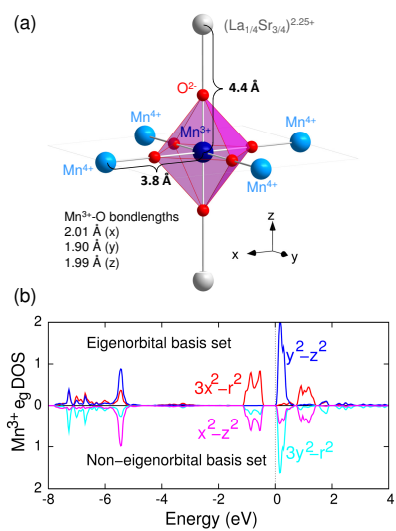}
  \caption{(Color online) (a) A sketch of the surrounding of a Mn$^{3+}$ ion
    in La$_{0.5}$Sr$_{1.5}$MnO$_{4}$ illustrates a different contribution of
    the further neighbors to the crystal field. The local Mn$^{3+}$-O bondlengths
    are also shown. (b) The Mn$^{3+}$ $e_g$ PDOS calculated by LSDA, projected onto the
    ($3x^2$--$r^2$,$y^2$--$z^2$) basis giving eigenstates (upper panel),
    and onto the ($x^2$--$z^2$,$3y^2$--$r^2$) basis, with mixed occupation
    of both orbitals (lower panel).}
    \label{Fig-crystal-field}
\vspace{-0.2cm}
\end{figure}

As mentioned above, the obtained orbital occupation contradicts
a common local JT picture. Two factors contribute to this. First,
there are longer-range contributions to crystal field, cf. e.g.
Refs~\cite{Pavarini_PRL_2004,Fang_PRB_2004}. Especially in this
layered manganite the coordinations of the Mn$^{3+}$ ion for
further in-plane neighbors are very different from the
out-of-plane ones. The further in-plane neighbors are the
Mn$^{4+}$ ions, which have higher valencies and are at
shorter distances than the further out-of-plane neighbors, being
the (Sr$^{2+}$, La$^{3+}$) ions, see Fig.~\ref{Fig-crystal-field}(a).
This has important consequences for the competition between
the rod-like $3x^2$--$r^2$ and the cross-like
$x^2$--$z^2$ to become the lowest in energy. Our
LSDA calculations reveal that the rod-like $3x^2$--$r^2$ orbital
lies 90 meV lower than the cross-like $x^2$--$z^2$ orbital. As a
consequence, the former becomes an occupied eigen-orbital~\cite{Brink_PRL_1999},
while the latter is not, i.e. the occupied
$e_g$ state of the Mn$^{3+}$ ion 
is a mixture of $x^2$--$z^2$ and $3y^2$--$r^2$ if one chooses
these orbitals as basis, see Fig.~\ref{Fig-crystal-field}(b). A
possible second factor is that the rod-like orbitals are more
in-plane, thereby maximizing the gain in kinetic energy in
this layered system. This factor is especially important for the
systems close to the localized-itinerant crossover, to which our
system belongs. The importance of these two factors is confirmed by
recent model calculations~\cite{Sboicleakov-to-be-published}.

\begin{figure}
 \includegraphics[angle=270,width=6cm]{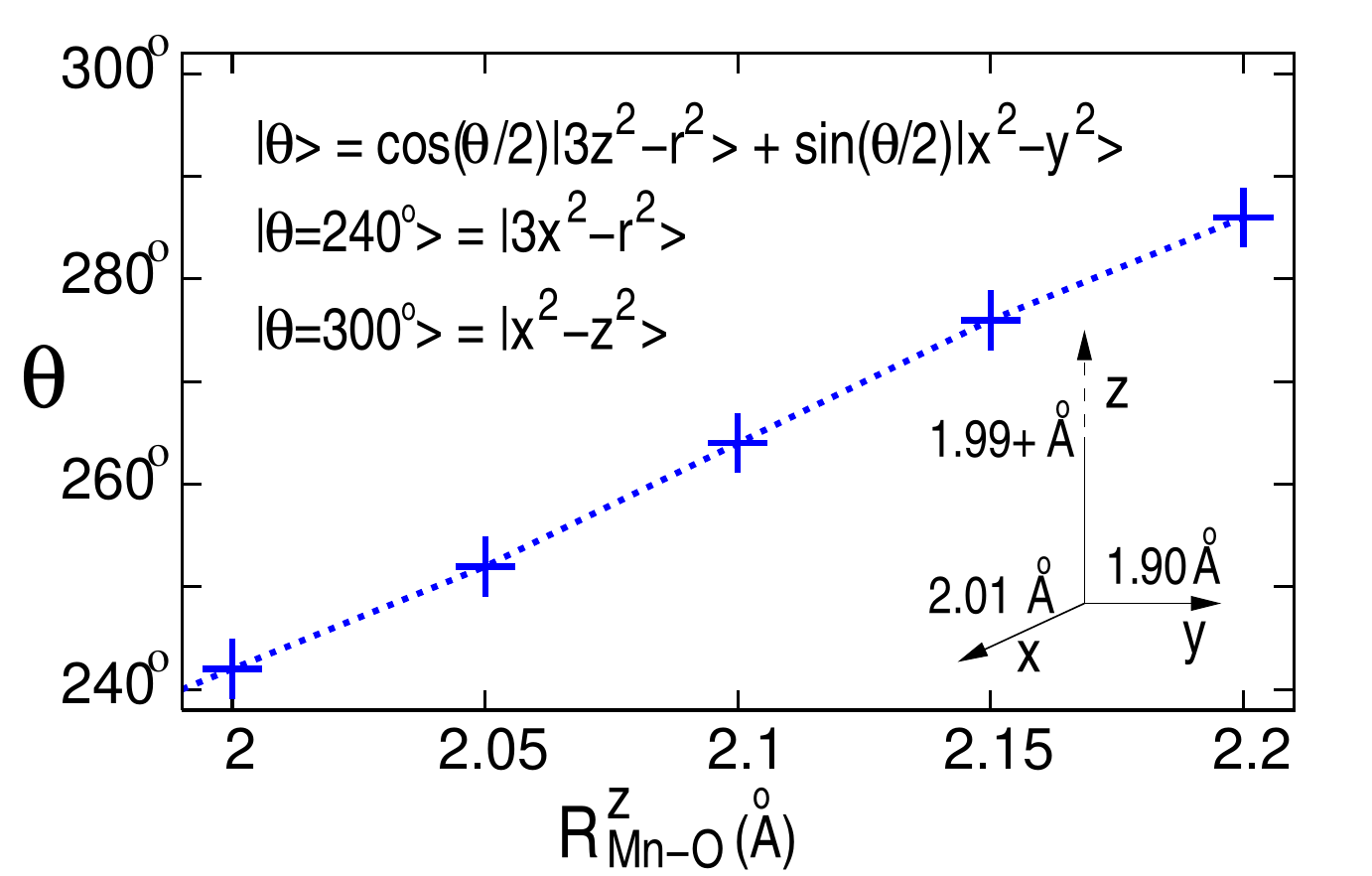}
  \caption{(Color online) Variation of the Mn$^{3+}$ $e^1_g$
    orbital state with respect to the $c$-axis Mn-O bondlength $R^{\rm z}$, in terms of the
    orbital mixing angle $\theta$ calculated by LSDA+$U$ ($U$=5 eV). The error bar is
    $\pm$4$^{\circ}$ for $U$ = 0-8 eV. The insets show
    the definition of the orbital states, and the real Mn$^{3+}$-O bondlengths
    with $R^{\rm z}$= 1.99 \AA~ at which the OO state is practically the pure
    $3x^2$--$r^2$/$3y^2$--$r^2$ type as discussed in the main text.}
    \label{Fig-Mn3+_wf}
\vspace{-0.2cm}
\end{figure}

Nevertheless, if the $c$-axis Mn$^{3+}$-O bondlength
were increased still further, the Mn$^{3+}$ $e^1_g$ state would
start to deviate from the $3x^2$--$r^2$ state, 
as seen in Fig.~\ref{Fig-Mn3+_wf}. If that bondlength exceeded 2.2 \AA, the
$e^1_g$ state would become more like $x^2$--$z^2$. Note, however,
that the actual stretching of the MnO$_6$ octahedra in {\LSMO}
along the $c$-direction, 1.99 \AA~, caused by the layered structure itself,
is much less than this 2.2 \AA~ value~\cite{Struct_opt}.

To summarize, using the linear dichroism and cluster model simulations of 
soft x-ray absorption spectroscopy, and density
functional calculations, we demonstrate that
La$_{0.5}$Sr$_{1.5}$MnO$_{4}$ displays the rod-like
$3x^2$--$r^2$/$3y^2$--$r^2$ orbital ordering at Mn$^{3+}$ sites.
Contrary to a common local JT picture, the observed orbital occupation 
strongly deviates from the one
expected from the local distortion of MnO$_6$ octahedra, i.e. the
assumption that the orbital and distortion mixing angles are the
same, $\theta$ = $\theta'$, is violated. We explain this by
considering the contribution of further neighbors to the
crystal-field splitting, and by the important role of the electron
kinetic energy in a system close to an insulator-metal transition.
We expect that our findings are not only applicable for the `214'
oxides, but also for layered materials in general, including the
highly topical artificial multilayer oxide systems.

This work is funded by the DFG via SFB 608.

\end{document}